\documentstyle[12pt]{article}
\begin{document}
\setlength{\unitlength}{1mm}
{\hfill  JINR E2-93-371, October 1993 } \vspace*{2cm} \\
\begin{center}
{\Large\bf On Quantum Deformation Of The Schwarzschild Solution}
\end{center}
\begin{center}
{\large\bf D.I.Kazakov$^{\ast}$, S.N.Solodukhin$^{\ast \ast}$}
\end{center}
\begin{center}
{\bf Laboratory of Theoretical Physics, Joint Institute for
Nuclear Research, Head Post Office, P.O.Box 79, Moscow, Russia}
\end{center}
\vspace*{2cm}
\abstract

We consider the deformation of the Schwarzschild solution in general
relativity due to  spherically symmetric quantum fluctuations of the
metric and the matter fields. In this case, the 4D theory of gravity
with Einstein action reduces to the effective two-dimensional dilaton
gravity. We have found that the Schwarzschild singularity at $r=0$ is
shifted to the finite radius  $r_{min} \sim r_{Pl}$, where the scalar
curvature is finite, so that the space-time looks regular and consists
of two asymptotically flat sheets glued at the hypersurface of constant
radius.
\vskip 4cm
\noindent $^{\ast}$ e-mail: kazakovD@thsun1.jinr.dubna.su  \\
\noindent $^{\ast \ast}$ e-mail: solodukhin@main1.jinr.dubna.su
\newpage
\section{Introduction} \setcounter{equation}0

   One of the most important unresolved problems in general relativity
is the problem of singularities. According to the results of Penrose and
Hawking [1], the space-time singularities are typical for a classical theory
of gravitation. Under rather general assumptions about the properties of the
matter they occur in the Universe and inside  black holes. The
curvature of space-time increases without limit near a singularity. In
such  circumstances the classical theory is not applicable and, in
particular, we cannot believe in its predictions concerning the
complete global structure of space-time.

   On the other hand, it is commonly believed that a successful
quantization of gravity will provide us with modifications to the
theory which are necessary to avoid the prediction of  geodesically
incomplete space-time manifolds [2,3].   Quantum corrections may
completely change the gravitational equations and the corresponding
space-time geometry at the Planck scale. The main problem on this way
is the non-renormalizability of the Einstein gravity since the
straightforward exploiting of the standard perturbation methods leads
to inconsistent quantum theory. However, quantum gravity can be
treated semi-classically [4] and  the obtained results are sensible in
some regimes when part of gravitational degrees of freedom in
the leading order can be considered as classical [5], while the other
part is described by exactly solvable quantum theory.

   In this paper, we are trying to take into account the influence of
quantum corrections on the behaviour of the Schwarzschild solution.
This solution is probably the most important one in general relativity.
It describes the space-time outside the gravitating body of mass $M$
and allows maximal Kruskal extension which has a singularity at the radial
parameter $r=0$.

Our strategy is the following.
We are interested in  spherically symmetric solution of
gravitational field equations and its deformation due to quantum
excitations of the metric and  the matter fields.  Therefore, it is
naturaly to assume that the general element $g_{\mu \nu}$ of the
space of all metrics ( over which we have to integrate in
functional integral) in the neighbourhood of the classical configuration
is represented as a sum of a spherically-symmetric part $g^{sph}_{\mu
\nu}$ and a non-spherically symmetric perturbation $h_{\mu \nu}$:
\begin{equation} g_{\mu \nu} = g^{sph}_{\mu \nu} + h _{\mu \nu} .
\end{equation}
We do not assume the spherically symmetric part
$g^{sph}_{\mu \nu}$ to be small and will quantize it exactly. Instead
a non-spherically symmetric deviation $h_{\mu \nu}$ is assumed to be
small and we will take it into account  perturbatively under quantization.
Spherically symmetric excitations of the metric do not contain
propagating modes while the modes of $h_{\mu \nu}$ do propagate.  So
in the first order, eq.(1.1) is a separation on  propagating and
non-propagating modes. Bearing in  mind the non-renormalizability of
quantum gravity with the Einstein-Hilbert action
\begin{equation}
S_{gr}[g_{\mu\nu}]=-{1 \over {16 \pi \kappa}} \int\limits_{}^{}d^{4}x
\sqrt{-g} R \end{equation}
one can expect that it is related  to the
contribution of the propagating $h$-modes (gravitons) to the Feynman
diagrams. Thus, quantum theory  blows up already in the first non-trivial
order in $h$.

   Inserting (1.1) into the Einstein-Hilbert action (1.2) we get in the
second order in $h$ :
\begin{equation}
S_{gr}[g_{\mu\nu}]=S_{gr}[g^{sph}_{\mu\nu}] + \int h \hat{D} h,
\end{equation}
where $\hat{D}$ is the second order differential
 operator determined with respect to the metric
$g_{\mu\nu}^{sph}$.\footnote{Actually, the $h$-modes can be expanded with
respect to the basis of
spherical harmonics and be represented as an infinite set of
two-dimensional fields (functions of only time and radial coordinates
$(t,r)$ ). The operator $\hat{D}$ is then the corresponding
two-dimensional second order operator.
}
To the leading order (which we only
will consider here) the non-spherical excitations $h_{\mu\nu}$ can be
considered as classical and consequently we can assume that
$h_{\mu\nu}=0$ in  (1.3).  So far, as  non-spherical excitations $h$
are concerned, we are in the classical regime. In this case, the  4D
theory of gravity with the Einstein-Hilbert action (1.2) reduces to
quantum theory of only spherical excitations of the metric with
the action $S_{gr}[g^{sph}_{\mu\nu}]$ describing  effective exactly
solvable two-dimensional theory  of the 2D dilaton gravity. In the
leading order, the effective theory describes the non-propagating
spherically symmetric modes and is correctly tractable under  the
quantization (in the sense of generalized renormalizability). It is possible
to compute the higher order corrections due to the presence of propagating
gravitons, although at some point one is bound to encounter the problem of
non-renormalizability of quantum gravity. The still unknown correct theory of
quantum gravity is likely to avoid this problem but we expect that the
modifications (which are not correctly calculable at the present moment)
will not drastically alter the leading order result.

   The 2D dilaton gravity has widely been investigated  recently
[6-14].  The review with detailed references can be found in [7]. The
4D dimensionally reduced models are discussed in [8,9].

\section{Effective two-dimensional theory}
\setcounter{equation}0

   The classical dynamics of a gravitational field interacting with the
matter is determined by the standard Einstein-Hilbert action
\begin{equation}
S=\int\limits_{}^{}d^{4}x \sqrt{-g^{(4)}} (-{1 \over 16 \pi \kappa}
R^{(4)} + {\cal L}_{mat}),
\end{equation}
where $R^{(4)}$ is the scalar curvature determined by a
four-dimensional metric $g^{(4)}_{\mu \nu}$ and $L_{mat}$ is the
 Lagrangian of matter fields.  The gravitational constant $\kappa$ has the
dimensionality of length squared $[l^{2}]$.

The metric $g^{(4)}_{\mu \nu}$ and the matter fields  are assumed
to be consistent with the condition of spherical symmetry. Let us
consider in detail the gravitational part of the action (2.1). An
arbitrary spherically symmetric metric can be written in the form
\begin{equation}
ds^{2}= g_{\alpha \beta}(z) dz^{\alpha}dz^{\beta} - r^{2}(z) (d\theta^2
+ \sin^{2} \theta d\varphi^2 ) ,
\end{equation}
where we assume that four-dimensional space-time is covered by
the coordinates $(z^{0}, z^{1}, \theta, \varphi)$.  Note that
$g_{\alpha \beta}$ in (2.2) plays the role of a metric on the 2D
space-time covered by the coordinates $(z^{0}, z^{1})$ and  $r^{2}(z)$
is a function on this two-dimensional space.\footnote{We will use
letters  $\mu, \nu =0,1,2,3,4$ for curved
indices in four dimensions, while the corresponding
indices in two dimensions will be denoted by $\alpha, \beta =0,1$}

   In a usual way we can choose the coordinates $(z^{+}, z^{-})$  in
which the first term in (2.2) takes the conformally flat form
\begin{equation}
ds^{2}s= e^{ \sigma(z^{+},z^{-})} dz^{+}dz^{-} - r^{2}(z^{+}, z^{-})
(d\theta^2 + \sin^{2} \theta d\varphi^2)  .
\end{equation}
In the case of the Schwarzschild metric, $(z^{+}, z^{-})$ are the
Kruskal coordinates defining the maximal extension of the black hole
space-time and "$r$" is indeed the radius measured from the singularity
located at the point $r=0$.

   The non-zero  Ricci tensor components for the metric (2.3) are the
following:
\begin{eqnarray} &&R_{+-}= \partial_{+} \partial_{-} \sigma +
\partial_{+} \partial_{-} \ln{r^{2}} + {1 \over 2} \partial_{+} \ln
r^{2} \partial_{-} \ln r^{2} , \nonumber \\ &&R_{\theta \theta}= -1-2
e^{-\sigma} \partial_{+}\partial_{-} r^{2} , \nonumber \\ &&R_{\phi
\phi}= \sin^{2} \theta R_{\theta \theta} , \nonumber \\ &&R_{ \pm \pm}=
\partial^{2}_{ \pm} \ln r^{2} + \partial_{ \pm} \ln r^{2}
\partial_{\pm} \ln r^{2} - \partial_{\pm} \sigma \partial_{\pm} \ln
r^{2}, \end{eqnarray}
and the scalar curvature $R^{(4)}=
4R_{+-}e^{-\sigma} -{2 \over r^{2}} R_{\theta\theta}$ is:
\begin{equation}
R^{(4)}= 4e^{-\sigma} \partial_{+} \partial_{-} \sigma - 2 e^{-\sigma}
\partial_{+} \ln r^{2} \partial_{-} \ln r^{2} + {2 \over r^{2}}
+ {8 \over r^{2}} e^{-\sigma} \partial_{+}\partial_{-} r^{2}
\end{equation}
Note that the first term in (2.5) coincides with the scalar curvature
$R^{(2)}$ of the two-dimensional metric $g_{+-}=e^{\sigma(z^{+},
z^{-})}$:  $R^{(2)}= 4e^{-\sigma} \partial_{+}\partial_{-} \sigma$.
Analogously, the whole expression (2.5) can be written in the
covariant form with respect to the two-dimensional metric $ds^{2}=
g_{\alpha\beta}dz^{\alpha}dz^{\beta}$
\begin{equation} R^{(4)}=
R^{(2)} - {2 \over r^{2}} ( \nabla r)^{2} + {2 \over r^{2}} + {2 \over
r^{2}} \Box r^{2},  \end{equation}
where $ \Box = \nabla^{2} = g^{\alpha\beta}\nabla_{\alpha}\nabla_{\beta}$.

   Since the determinant of the metric (2.3) is $g^{(4)}= -{1 \over 4}
e^{2\sigma} r^{4} \sin^{2}\theta$ , the gravitational action
\begin{equation} S_{gr}=-{1 \over 16 \pi \kappa}
\int\limits_{}^{}d^{2}z \int\limits_{0}^{2\pi} d\varphi
\int\limits_{0}^{\pi}d\theta \sqrt{-g^{(4)}} R^{(4)} \end{equation}
takes the form:
\begin{equation}
S_{gr}=- {1 \over 8\kappa}\int\limits_{}^{}d^{2}z[r^{2}R^{(2)} -2(\nabla r)^{2}
+2] ,
\end{equation}
where we have omitted the integral of $\Box r^{2}$ which is the total
derivative and does not affect the equations of motion.  The action
(2.8) determines the dynamics of  spherically symmetric excitations
of the 4D gravitational field. On the other hand, (2.8) describes the
effective two-dimensional scalar-tensor theory of gravity.  It is worth
noting that this theory is indeed of the 2D dilaton gravity type. It is
easy to see this introducing the "dilaton" field $\phi=\ln
({r^{2} \over \kappa})$. Then eq.(2.8) takes form of the dilaton gravity
[6-7]:
\begin{equation} S_{gr}=-{1 \over
8}\int\limits_{}^{}d^{2}z[e^{\phi} (R - {1 \over 2} (\nabla \phi)^ {2})
+ U(\phi)] , \end{equation}
where the "dilaton" potential is $U(\phi)={2 \over
\kappa}$. This observation is important for us since it allows one
to use all the methods previously developed for the 2D dilaton gravity
[6-7,12-14]. Note that usually one considers the following
dilaton-gravity action \begin{equation} S_{str}=-{1 \over
4}\int\limits_{}^{}d^{2}z \sqrt{-g} e^{\phi}[R-(\nabla \phi)^{2} +
\lambda] ,\end{equation}
which is inspired by string models. The
essential difference between (2.9) and (2.10) lies in the quantum
region. The string-inspired action (2.10) is shown to be finite, while
the action (2.9) is  renormalizable in the generalized sense:
quantum corrections change  the form of the potential
$U(\phi)$.  Having this in  mind, let us consider instead of (2.8)
the generalized action with an arbitrary "dilaton" potential $U(r)$:
\begin{equation} S_{gr}=-{1 \over 8} \int\limits_{}^{}d^{2}z
\sqrt{-g}[r^{2}R^{(2)}-2 (\nabla r)^{2} + {2 \over \kappa} U(r)] ,
\end{equation} where we have introduced  the dimensionless variable
$r \rightarrow {r \over \sqrt{\kappa}}$.  Varying (2.11) with respect
to the 2D metric $g_{\alpha\beta}$ and $r$  leads to the equations of
motion
\begin{eqnarray}
&&2r\nabla_{\alpha}\nabla_{\beta}r=g_{\alpha\beta}[{1 \over \kappa} U(r) + 2r
\Box r +(\nabla r)^{2}] , \nonumber \\
&&2\Box r + rR + {1 \over \kappa}U^{'}(r)=0  .
\end{eqnarray}
The first equation in (2.12) means the existence of the two-dimensional
Killing vector [10-12] $\xi_{\alpha}= \varepsilon_{\alpha}^{\ \beta}
\partial_{\beta}r$ $(\nabla_{\alpha} \xi_{\beta} + \nabla_{\beta}
\xi_{\alpha}=0)$. Thus, we can choose the field $r$ as one of   the
coordinates (which is space-like) and use the Schwarzschild
gauge where the metric takes the form:  \begin{equation} ds^{2}=gdt^{2}
- g^{-1} \kappa dr^{2}, \end{equation} where $g=g(r)$.
For the metric
(2.13) we get  $\Box r=-{g^{'} \over \kappa}$, $ (\nabla
r)^{2}=- {g \over \kappa}$. Consequently,  one has the following
solution of equations (2.12):
\begin{equation} g(r)=- {2M \over r} +
{1 \over r} \int\limits_{}^{r}U(\rho) d\rho , \end{equation}
where $M=const$. If $U(r)=1$, then $g(r)=1- {2M \over r}$ and we obtain the
Schwarzschild metric. It is not surprising since the Schwarzschild
metric is the unique spherically symmetric solution of Einstein
equations in empty space.  The constant $M$ coincides with the ADM mass
calculable at space-like infinity.  It is worth noting that for $U=1$
eqs.(2.12) are the  Einstein equations in empty space
\begin{equation}
G_{\mu\nu} \equiv R_{\mu\nu}^{(4)} - {1 \over 2} g^{(4)}_{\mu\nu}
R^{(4)}=0  \end{equation} considered on the spherical metric (2.2-3).
Hence, the reduction to the effective 2D theory (2.8) is self-consistent
and we obtain again the Einstein equations.

  To complete our consideration we present here the expressions
for the effective two-dimensional and four-dimensional  scalar
curvature valid for the metric (2.2), (2.13-14):
\begin{equation} R^{(2)}= -g^{''}=- {2M \over r^{3}} + {2 \over \kappa
r^{3}} \int\limits_{}^{r} U(\rho)d\rho -{2 \over \kappa r^{2}}U +
{U^{'} \over {\kappa r }} \end{equation}
\begin{equation} R^{(4)}= {2 \over \kappa r^{2}} (1-U(r))-{U^{'} \over
\kappa r} \end{equation}
Note that for $U=1$ we get $R^{(4)}=0$
everywhere in the external region of gravitating body, as it follows
from eqs.(2.15). In the next sections, we will show that taking into
account  quantum spherically symmetric excitations leads to the
deformation of the form of the potential $U(r)$. According to (2.17), it
 manifests itself in non-zero value of $R^{(4)}$ outside the
gravitating body.  The Einstein tensor $G_{\mu\nu}$(2.15) for the
metric (2.2) can also be written covariantly with respect  to
the 2D metric $g_{\alpha\beta}$:
\begin{eqnarray} &&G_{\alpha\beta}={2
   \over r} \nabla_{\alpha} \nabla_{\beta} r -g_{\alpha\beta} ({1 \over
r^{2}} +{2 \over r} \Box r + {(\nabla r)^{2} \over r^{2}}), \ \alpha ,
\beta =0,1 ,\nonumber \\ &&G_{\theta\theta}={r^{2} \over 2} (R^{(2)} +
{2 \over r} \Box r) ,\nonumber \\ &&G_{\varphi\varphi}=\sin^{2}\theta
G_{\theta\theta} .\end{eqnarray}
Eqs.(2.12) look like the quantum-corrected Einstein equations
  \begin{equation}
G_{\mu\nu}=T^{eff}_{\mu\nu}, \end{equation} where
\begin{eqnarray}
&&T^{eff}_{\alpha\beta}={g_{\alpha\beta} \over r^{2}}[U( {r \over
   \sqrt{\kappa}})-1],
\nonumber \\
&&T^{eff}_{\theta\theta}=-{1 \over 2\kappa} r \partial_{r} U({r \over
   \sqrt{\kappa}}) ,
\nonumber \\
&&T^{eff}_{\varphi\varphi}= \sin^{2} \theta T_{\theta\theta}
\end{eqnarray}
is the effective energy-momentum tensor which is induced
due to  quantum spherically symmetric excitations.

\bigskip

\section{Quantization of spherically-symmetric \protect \\
excitations} \setcounter{equation}0
Consider now  quantum theory of spherically symmetric
excitations of the metric described by the two-dimensional effective
theory with the action (2.8):
\begin{equation} S_{gr}=-{1 \over 8}
\int\limits_{}^{}d^{2}z \sqrt{-g}[r^{2}R^{(2)}-2 (\nabla r)^{2} + {2
\over \kappa} ] , \end{equation} where $r$ is dimensionless.  Note that
the dimensional gravitational constant  $\kappa$ from the combination
${1 \over \kappa} R^{(4)}$ in (2.1) has moved to the $\lambda$-term in the
2D action (3.1). It reflects the fact that the 2D effective theory
(3.1) has a better renormalizable property than the initial 4D-action
(2.1).  As it has been noted in Sect.2, the action (3.1) takes the form of the
2D
dilaton gravity which is widely investigated in recent years in
connection with the interest in  two-dimensional black holes
[6-7].  In particular, the theory (3.1) was shown to be generally
renormalizable in the sense  that the renormalized action takes the same
form as the original action (3.1) with some new potential $U(r)$:
\begin{equation} S_{gr}=-{1 \over 8} \int\limits_{}^{}d^{2}z
\sqrt{-g}[r^{2}R^{(2)}-2 (\nabla r)^{2} + {2 \over \kappa} U(r)] .
\end{equation}
In the conformal gauge $g_{\alpha\beta}=e^{2\sigma}
\bar{g}_{\alpha\beta}$ the action (3.2) takes the form
\begin{equation} S_{gr}= \int d^{2}z \sqrt{-\bar{g}} [ {1 \over 2 \psi}
\bar{g}^{\alpha\beta} \partial_{\alpha}\psi \partial_{\beta} \psi + 2
\bar{g}^{\alpha\beta} \partial_{\alpha} \psi \partial_{\beta} \sigma -
\psi \bar{R} -{1 \over 4\kappa} U(\psi) e^{2\sigma}] ,\end{equation}
where we denoted $\psi= {r^{2} \over 8}$ and $\bar{R}=R^{(2)}[\bar{g}]$.

Let us start with neglecting a possible anomalous term in the quantum
version of the action (3.3). Of course, it is just an approximation, but its
advantage is that the results obtained have a very simple analytic form and,
moreover, possess the same interesting properties as in a more
general case.

   The divergencies now can be calculated by the background-field method
[12]. It is useful to interpret (3.3) as the $D=2$ $\sigma$-model\footnotemark\
\addtocounter{footnote}{0}\footnotetext{Note that our definition of curvature
differs
in sign from that of the paper [12].}:
\begin{eqnarray}
&&S=\int d^{2}z \sqrt{-\bar{g}}[{1 \over 2} G_{ij}(X) \bar{g}^{\alpha\beta}
\partial_{\alpha}X^{i}\partial_{\beta} X^{j} -{1 \over 2}\bar{R} \Phi (X) +
T(X)], \nonumber \\
&&X^{i}=(\psi, \sigma); \Phi =2\psi; \ T=- {1 \over 4\kappa} U(\psi)
e^{2\sigma};
\nonumber \\
&&G_{ij}=\left(\matrix {
            {1 \over \psi}  &2\cr
             2              &0
   \cr}\right)    .
\end{eqnarray}

Since the metric $G_{ij}$ is flat and the dilaton $\Phi$ is a linear
function, the unique nontrivial divergency corresponds to the
renormalization of the tachyon $T$. The coefficients of Weyl anomaly
corresponding to the tachyon coupling take the form [12]:
\begin{equation} \bar{\beta}^{T}=- G^{ij} \nabla_{i} \nabla_{j} T +
(G^{ij} \partial_{i} \Phi \partial_{j} T -2T) \equiv \beta^{T} + \Delta
\beta^{T} \end{equation} For (3.4) one gets \begin{equation} \Delta
\beta^{T}=0 , \ \beta^{T}=-{1 \over 4\kappa}e^{2\sigma}({1 \over \psi}
U - 2 \partial_{\psi} U) \end{equation}
Due to the factorization of the tachyon
$T$ (3.4) one obtains from (3.6) the $\beta$- function for the dilaton
potential $U(\psi)$:
\begin{equation} \beta^{U}=( {1 \over \psi} U - 2
\partial_{\psi} U) .\end{equation}
The fixed point of (zero of the
$\beta^{U}$-function) corresponds to the potential
\begin{equation}
U(\psi)=c \sqrt{\psi}= \tilde{c} r .
\end{equation}
In this case, the theory is finite [12] with the  potential
corresponding to the string-inspired dilaton gravity (2.10). A
weaker renormalization condition is satisfied if $( {1 \over \psi } U
-2\partial_{\psi} U)$ is proportional to the potential itself, i.e., in
the case of a Yukawa-like potential
\begin{equation}
U(r)= \epsilon
e^{-\lambda r^{2}} r .
\end{equation}
Then, the divergency can be
absorbed into a renormalization of $\epsilon$.  Inserting potentials
(3.8) or (3.9) into (2.14) we obtain a metric corresponding to the
UV fixed point. However, there exists a problem of reaching the fixed
point since the classical ("bare") potential $U(r)=1$ can be out of the
attraction region of the UV fixed point. Therefore  we have to consider
the renormalization group equation for the potential $U$:
\begin{equation} \partial_{t} U= \beta^{U}={1 \over \psi} U -
2 \partial_{\psi} U ,
\end{equation}
where $t= \ln { \mu \over \mu_{0}}$, $\mu$ being a scale parameter.

One should add
the "initial" condition to  eq.(3.10). We will assume that at $\mu = \mu_{0}$
($t=0$) the potential $ U( \psi, t) $ coincides  with the bare potential
\begin{equation}
U(\psi, t=0)=1 .
\end{equation}
It is easy to find a general solution of eq.(3.10):
\begin{equation}
U(\psi, t)=\sqrt{\psi} f(\psi -2t),
\end{equation}
where $f(...)$ is still an arbitrary function to be chosen
from the initial condition.

Note that in our case   eq.(3.10) is considered
in the region $\{t \geq 0, \psi \geq 0 \}$ being the transport equation
with the characteristic line $\psi - 2t=0$.  Therefore, the
solution $U$ at the point $(\psi, t)$ below this line ( $\psi > 2t$) is
obtained by transporting   the initial
condition, in our case it is $U(\psi, t=0)=1$, along the characteristic.
Hence, for $\psi >2t$ we
get from (3.11-12) that $f(\psi) = {1 \over {\psi}^{1/2} }$ and
consequently
\begin{equation} U(\psi, t)={ \psi^{1/2} \over (\psi
-2t)^{1/2} }, \ \ \ \psi >2t .\end{equation}

On the other hand, above the line $\psi =2t$ ( t.e. for $\psi < 2t$ )
the boundary condition $U(\psi,t)|_{\psi=0} = u(t)$ is "transported".
We do not have this kind of a boundary condition from our problem.
However, as one can see from the form of the general solution (3.12),
the value of $U(\psi, t)$ for $\psi=0$ cannot be
different from zero. In other words, there is the unique possible
boundary condition \begin{equation} U(\psi, t)|_{\psi=0}=0
\end{equation} consistent with eq.(3.10).  Thus, the solution of (3.10)
in the region $\{ \psi \geq 0, t \geq 0 \}$ takes the following form:
\begin{eqnarray}
U(\psi, t)= \left\{
\begin{array}{ll}
        0     & {\rm if\ } \psi \leq 2t , \\
{ \psi^{1/2} \over ( \psi -2t)^{1/2} }  & {\rm if\ } \psi > 2t .
\end{array}
\right.
\end{eqnarray}
Note that the function (3.15)  has a discontinuity along the
characteristic line $\psi=2t$. Remembering that $\psi = {r^{2} \over 8
\kappa }$ we obtain

\begin{eqnarray}
U(r, t)= \left\{
\begin{array}{ll}
        0     & {\rm if\ }  0 \leq r \leq 4 \sqrt{\kappa t} , \\
{ r \over ( r^{2} - 16 \kappa t)^{1/2} }  & {\rm if\ } r > 4
\sqrt{\kappa t}.  \end{array} \right.  \end{eqnarray}
As we expected, the
fixed point (3.8) (or (3.9)) is not reached in the limit $t \rightarrow
+ \infty$.  However, the t-dependence of the solution (3.16) can be absorbed
into the redefinition of the gravitational constant $\kappa$:
$\kappa^{*} = \kappa t$. Then, (3.16) can be written in the form:
\begin{eqnarray} U(r,\kappa^{*})= \left\{ \begin{array}{ll} 0     &
        {\rm if\ } 0 \leq r \leq  4 \sqrt{\kappa^{*}}, \\ { r \over (
r^{2} - 16 \kappa^{*} )^{1/2} }  & {\rm if\ } r > 4 \kappa^{*} .
\end{array}
\right.
\end{eqnarray}
We will omit $*$ later on.

Let us now return to eq.(2.14) connecting the metric of the spherically
symmetric solution with the potential $U(r)$. We get that the quantum
spherically symmetric gravitational fluctuations lead to the following
deformation of the Schwarzschild metric:
\begin{equation} g(r)=- {2M
\over r} + {1 \over r} (r^{2} - a^{2})^{1/2}, \end{equation} where $r
\geq a=4 \sqrt{\kappa}$.

Analyzing the metric (3.18) we discover that the singularity of the
Schwarzschild solution at $r=0$  is now shifted to the finite radius
$r=a$  (in four-dimensional picture it means that singularity now is
"spread" over a two-dimensional sphere of radius $r=a$). One can see
this when calculating the 4D scalar curvature (2.17)
\begin{equation} R^{(4)} = { 1 \over a^{2} } [ 2 x^{2} (1 - { 1 \over
\sqrt{1- x^{2}} } ) + x^{4} (1- x^{2} )^{-3 /2} ], \end{equation}
where $x={a \over r}$. In the limit $x \rightarrow 1$ we have $R^{(4)}
\rightarrow + \infty $. On the other hand, for large $r$ ( $x
\rightarrow 0$) we obtain
\begin{equation} R^{(4)} (r) \approx {2
\over r^{2} } ({ a \over r})^{4} .\end{equation}

One can also get from (3.18) the asymptotic expression for the metric
$g(r)$ for large $r >> a$:  \begin{equation} g(r) \approx 1 - {2M \over
r} - {a^{2} \over {2r^{2}}} .\end{equation}

Eq.(3.21) looks like the metric of a charged body with the mass $M$ and
the charge $Q^2=a^2/2$ but with opposite sign in front of
the charge's term. It is interesting that the potential $U$ in (2.14),
(3.18) does not lead to an additional contribution to the mass $M$,which
is due to the fact that the $U$-term in (2.14), (3.18) has the
asymptote $\sim { 1 \over r^{2}}$ for large $r$.

It is worth noting that as follows from the general expression (2.17)
(and (3.19)) the scalar curvature $R^{(4)}(r)$ does not depend on
 the mass of a gravitating body, it is rather universal and is
determined by the parameters of the gravitational field itself (via the
gravitational constant).  Later on, the minimal radius $a$ will be
supposed to be equal to the Planck radius $r_{pl}$ by assuming that
possible factors $\sim 10$ are non-essential.

We see from (3.18) that the Schwarzschild horizon at $r_{h}=2M$ is
also shifted $r_{h}= \sqrt{4M^{2} + a^{2}}$ so that the asymptotically flat
metric
(3.18) describes the space-time with the same causal structure as the
Schwarzshild one. At the same time, the Schwarzshild singularity at $r=0$
manifests itself both in the metric and in the curvature ($R^{(4)} \sim
\delta (r)$) while the deformed metric (3.18) is regular at $r=a$ and
only the scalar curvature $R^{(4)}(r)$ is still singular.

Formally, there exists an extension of the metric (3.18) behind the
singularity at $r=a$. To see this, let us change the variable: $r=a
\cosh{x}$.  Then, the space-time for $x>0$ has the metric (3.18) while
for $x<0$ the metric takes the same form (3.18) but with sign $(-)$ in
front of the second term. In terms of the variable $r$ it simply means that
eqs.(3.17)-(3.18) are extended to the other branch of the square root
function. So we obtain that the resulting metric is a two-valued function of
the radius $r$:

\begin{equation}
g^{(\pm)}= -{2M \over r} \pm {1 \over r} (r^{2} - a^{2}) ^{1/2}
\end{equation}
At the point $r=a$,  both functions $g^{(+)}(r)$ and $g^{(-)}(r)$
(but not their derivatives) are glued continuously.  The
scalar curvature on the $(-)$-sheet takes the form
\begin{equation}
R^{(4)}_{(-)} = { 1 \over a^{2} } [ 2
x^{2} (1 + { 1 \over \sqrt{1- x^{2}} } ) - x^{4} (1- x^{2} )^{-3 /2} ],
\end{equation}
where $x={ a \over r}$. In the limit $x\rightarrow  1$ we have $R^{(4)}_{(-)}
\rightarrow - \infty$. The $(-)$-sheet is also asymptotically flat but for
large $r$  $R^{(4)}_{(-)}(r)$ has a different type of asymptote than for
the $(+)$-sheet (3.20):
\begin{equation}
R^{(4)}_{(-)}(r) \approx {4 \over r^{2}}.
\end{equation}
Note that for positive $M$ the function $g_{(-)}(r)$ is negative
everywhere. So the whole $(-)$-sheet is behind the horizon $(r=r_{h})$
from the point of view  of an observer staying on the $(+)$-sheet.
At this stage such a picture of complete space-time seems to be
formal since no observer can penetrate through  the
singularity at $r=a$ and appear in the $(-)$-sheet.  However, we will
see  in the next section that taking into account the conformal anomaly
we obtain the space-time with the same structure but
with the smooth behaviour at $r=a$. The extended space-time
then occurs to be regular everywhere  and geodesically complete.

\bigskip

\section{Solution with the anomaly}
\setcounter{equation}0

Let us consider now the deformation of the potential $U(r)$ due to
the fluctuations of the ghost and  matter fields.
Taking into account only the spherically symmetric excitations we
obtain that these fields contribute to the quantum effective action
via the Weyl anomaly from the gravitational integration measure and
that of matter fields.

In two-dimensional dilaton gravity there is a well known ambiguity [13]
in choosing the 2D metric to construct the Faddeev-Popov ghost
determinant.  One may use any metric of the form:$g_{\alpha}=r^{\alpha}
g$, where $r= \exp{\phi \over 2}$ is dilaton and $\alpha$  is
an arbitrary constant.  However, in our case the measure of integration
over the 2D gravitational fields $(r, g^{(2)}_{\alpha\beta})$ is
induced by integration over the 4D metric $g^{(4)}_{\mu\nu}$ (in
assumption of spherical symmetry) and consequently should be determined
with respect to the rescaled metric $\hat{g}_{\alpha\beta}=
r^{2}g_{\alpha\beta}$, so no ambiguity arises.

Hence,  after gauge fixing
 $g_{\alpha\beta}= e^{2\sigma} \bar{g}_{\alpha\beta}$, the action will
be supplied with the usual logarithm of the  Faddeev-Popov ghost
determinant [13]:
\begin{equation}
S_{FP}={24 \over 96 \pi}
\int\limits_{}^{}d^{2}z \sqrt{-\hat{g}} R_{\hat{g}} \Box^{-1}_{\hat{g}}
R_{\hat{g}},
\end{equation}
where $R_{\hat{g}}$ is the 2D scalar curvature
determined by the metric $\hat{g}_{\alpha\beta}$.

Considering the spherically symmetric configurations of matter
fields we find that they are described by some effective 2D action
which in general takes the form: ${\cal L}^{eff}_{mat}={\cal L}_{(k)}
r^{2k}$, where $k$ runs over positive and negative integers (for
example, the $k=1$ term appears for 4D scalar fields).  We will consider
here only the simplest case of decoupled dilaton $r$ when
${\cal L}_{(0)}=\sum_{i=1}^{N}(\nabla f^{i})(\nabla f^{i})$ is
the action for the 2D conformal fields. The integration measure for the
matter fields $f^{i}$ is determined with respect to non-rescaled metric
$g_{\alpha\beta}$.

Thus we come to the following quantum effective action:
\begin{eqnarray}
&&S=S_{gr} + S_{FP} +NS_{anom}, \nonumber \\
&&S_{anom}=- {1 \over 96\pi} \int\limits_{}^{}R^{(2)}_{g} \Box^{-1}_{g}
R^{(2)}_{g} .
\end{eqnarray}
Using the identity $R_{\hat{g}}= {1 \over r^{2}} \Box_{g} \ln r^{2} +
{1 \over r^{2}} R_{g}$, the action (4.2) can be rewritten in the
form
\begin{equation} S=S_{grav} - {A \over 8} \int\limits_{}^{}R_{g}
\Box^{-1}_{g}R_{g} + {B \over 2} \int\limits_{}^{}(R_{g} \ln r
-{(\nabla r)^{2} \over r^{2}}) \sqrt{-g} d^{2}z, \end{equation} where
$A={N-24 \over 12\pi}$, $B={24 \over 12\pi}$.

In the conformal gauge $g_{\alpha\beta}=e^{2\sigma} \bar{g}_{\alpha\beta}$ we
obtain from (4.3)
\begin{eqnarray}
&&S=\int\limits_{}^{}d^{2}z \sqrt{- \bar{g}} [ {1 \over 2\psi}(1 - {B \over
4\psi})(\nabla \psi)^{2} +2(1- {B \over 4\psi}) (\nabla \sigma)(\nabla \psi)
+ {A \over 2} (\nabla \sigma)^{2}  \nonumber \\
&&-{1 \over 2} \bar{R} (A\sigma +2\psi -{B \over 2} \ln \psi) -{1 \over
4\kappa}
U(\psi) e^{2\sigma} ] + S_{FP}[\bar{g}] + N S_{anom}[\bar{g}],
\end{eqnarray}
where $\psi={r^{2} \over 8\kappa}$ and $\bar{R}$ is the scalar
curvature corresponding to the metric $\bar{g}_{\alpha\beta}$.  The
action (4.4) again takes the form of the 2D $\sigma$-model (3.6) where
\begin{eqnarray}
&&X^{i}=(\psi, \sigma); \Phi (X) =2\psi + A\sigma -{B
\over 2} \ln \psi ; \ T=- {1 \over 4\kappa} U(\psi) e^{2\sigma};
\nonumber \\ &&G_{ij}=\left(\matrix { {1 \over \psi}(1-{B \over 4
\psi})  &2(1 -{B \over 4\psi})\cr 2(1-{B \over 4 \psi}) &A \cr}\right)
\end{eqnarray}
The metric $G_{ij}$ is flat and its determinant is
\begin{equation} det \ G =-4(1-{B \over 4\psi}) (1 - {(A+B) \over
4\psi}) .\end{equation}
Note that $A+B={N \over 12\pi} >0$ and $det \
G$ (4.6) is zero if $\psi={B \over 4}$ and $\psi={(A+B) \over 4}$. At
these points the kinetic term in (4.4) is singular and  if $A \neq 0$
changes its sign.

For the tachyon $\beta$-function we get
\begin{equation}
\Delta \beta^{T}=0, \ \beta^{T}=-G^{ij} \nabla_{i}\nabla_{j} T
\end{equation}
Analyzing this equation we consider  two different cases.

\bigskip

{\sl 4.1 A=0}

We begin with the consideration of the case when the matter fields
do not contribute to the effective action (4.3-4), i.e. $A=0$. One can
see that the target space metric then takes the form
\begin{equation}
ds^{2}_{targ}= (1-{B \over 4\psi}) ds^{2}_{B=0} ,
\end{equation}
where $ds^{2}_{B=0}= {1 \over \psi} d^{2} \psi + 4 d \psi d \sigma$
is the target metric of the $\sigma$-model (3.4). Due to the conformal
transformation of the 2D Laplacian, we get from (4.7) that
\begin{equation}
\beta^{T}= {1 \over {(1-{B \over 4 \psi})}} \beta^{T}_{B=0} ,
\end{equation}
where  $\beta^{T}_{B=0}$ is the beta function (3.6).

Thus, we obtain the renormalization group equation for the potential
$U(\psi)$
\begin{equation} \partial_{t} U= {1 \over (1-{B \over
4\psi})} [ {1 \over \psi} U- 2 \partial_{\psi} U] . \end{equation}
To solve this equation, let us introduce  the function $F(\psi)$  such
that
\begin{equation} dF=(1-{B \over 4\psi}) d \psi . \end{equation}
Choosing the integration constant so that $F(\psi={B \over 4})=0$
we obtain
\begin{equation}
F(\psi)=  \psi -{B \over 4} - {B \over 4} \ln {4\psi \over B} .
\end{equation}
As one can see, the second derivative $F''(\psi)$ (4.12) has
a discontinuity at the point $\psi ={B \over 4}$.

Eq.(4.10) is the transport equation with the characteristic
$F(\psi)-2t=0$.  Initially, the function $U(\psi,t)$ is defined in the
region  $\{ \psi >0, t \geq 0 \}$.  However, as one can see from the
form of the characteristic the line $\psi  =B/4$ is a singular one
since  the coefficient in front of  $\partial_{t} U$ in (4.10)
is zero. Therefore,  eq.(4.10) must be considered separately in
the regions $0<\psi \leq B/4$ and $\psi \geq B/4$. It is important that
the function $F(\psi)$ (4.12) has  single-valued inverse function
$F^{-1}(x)$ defined for $x \geq 0$.
The value of $U(\psi,t)$ for $ F(\psi)-2t >0$ is
obtained by transporting the  initial condition at $t=0$ along the
characteristic. On the other hand, the value of $U(\psi,t)$ for
$F(\psi)-2t <0$, is obtained by transporting the boundary condition:
$U|_{\psi={B/4}} =\mu(t)$. Considering eq.(4.10) in the
regions $0<\psi \leq B/4, \  \psi \geq B/4$ separately,  we
may choose different boundary conditions $\mu_{1}(t), \mu_{2}(t)$.
Actually, we have no concrete choice for $\mu(t)$. Therefore, for
simplicity we will assume that $\mu(t)=0$.  The general solution of
(4.10) is
\begin{equation} U(\psi, t)= \sqrt{\psi}
f(F(\psi)-2t) .\end{equation}
{}From the initial condition
\begin{equation}
U(\psi,t=0)=1
\end{equation}
we get the equation on the function $f$ in the region where
$F(\psi)-2t>0$
\begin{equation} f(F(\psi))=\psi^{- {1\over 2}}
\end{equation}
Thus, one obtains that $f(x)=(F^{-1}(x))^{-{1 \over 2}}$, where $F^{-1}(x)$ is
the function inverse to (4.12) single-valued in the regions
$0< \psi \geq B/4$ and $\psi \geq B/4$. Note that $F^{-1}(x)$ is a
continuous monotonic function but its derivative is
singular at  $x=0$.  Thus for $F(\psi)-2t >0$ we
have the following solution of (4.10) with the initial condition (4.14):

$$
U(\psi, t)= {\psi^{1 \over 2} \over (F^{-1}[F(\psi)-2t])^{1 \over 2}}
$$

On the other hand, for $F(\psi)-2t <0$ the form of the function $f(..)$
is defined by the boundary condition: $\sqrt{B/4} f(-2t)=\mu(t)$. Since we
have chosen $\mu(t)=0$, $f(..) \equiv 0$ in this region. Thus, the
complete solution of eq.(4.10) reads
\begin{eqnarray} U(\psi, t) =
\left\{ \begin{array}{ll} 0 & {\rm if\ } F(\psi)-2t <0 , \\ \psi^{1/2}
  \over (F^{-1}[F(\psi)-2t])^{1/2} & {\rm if\ }  F(\psi)-2t >0
\end{array}
\right.
\end{eqnarray}

Remembering now that $\psi={r^{2} \over 8\kappa}$ we express the
solution (4.16) in terms of the variables $(r,t)$. Note at first that
\begin{eqnarray}
&&F({r^{2} \over 8\kappa})={1 \over 8\kappa}F^{*}(r^{2}), \nonumber \\
&&F^{*}(r^{2})=
r^{2}-b^{2}-b^{2} \ln {r^{2} \over b^{2}} ,
\end{eqnarray}
where $b^{2}=2B\kappa$.
{}From (4.17) we have that $[F^{*}]^{-1} 8\kappa=8\kappa F^{-1}$
and consequently (4.16) takes the form
\begin{eqnarray}
U(r, t) = \left\{
\begin{array}{ll}
   0
  & {\rm if\ } F^{*}(r^{2}) <16\kappa t , \\
  r \over (F^{*-1}[F^{*}(r)-16\kappa t])^{1/2}
  & {\rm if\ }  F^{*}(r^{2}) >16 \kappa t .
\end{array}
\right.
\end{eqnarray}

As one can easily see  the $t$-dependence of the solution (4.18) again
can be absorbed into the redefinition of the constants
$\kappa^{*}=\kappa t, \ B^{*}={B \over t}$, with the
function $F^{*}$ being unchanged. Equation
$F^{*}(r^{2})-16 \kappa t=0$ has two roots:  $0<r_{1m}<b$ and
$r_{2m}>b$.  In terms of new variables the solution (4.18) looks  as
follows:
\begin{eqnarray} U(r) = \left\{ \begin{array}{ll} r \over
(F^{*-1}[F^{*}(r^{2}- 16\kappa])^{1/2} & {\rm if\ } 0<r \leq r_{1m} ,\\
 0 & {\rm if\ } r_{1m} \leq r \leq r_{2m} ,\\ r \over
   (F^{-1}[F(r^{2})-16 \kappa])^{1/2} & {\rm if\ }  r \geq r_{2m} .
  \end{array} \right.  \end{eqnarray}
The potential (4.19) tends  to (3.19) when $b \rightarrow 0$.

Let $r \geq r_{2m}$.
The potential $U(r)$ (4.19) is a continuous function in this region.
Near the point $r=r_{2m}$ it takes the form
\begin{equation}
U(r) \approx {r \over b} (1 -{1 \over \sqrt{2} b}
\sqrt{1-({b \over r_{2m}})^{2}}(| r^{2}- r^{2}_{2m}|)^{1/2}).
\end{equation}
Thus, at $r=r_{2m}$ it has a finite value: $U(r_{2m})= {r_{2m}
\over b}$.  However, the derivative of the potential is singular:
$\partial_{r} U(r) \rightarrow - \infty$.

The metric function $g= -{2M \over r} +{1 \over r}
\int\limits_{r_{2m}}^{r}U(\rho)d\rho$
in the vicinity of $r=r_{2m}$ can be written as follows:
\begin{equation}
g \approx -{2M \over r} + {r \over 2b} - {1 \over 3 \sqrt{2} b^{2}}
\sqrt{1-({b \over r_{2m}})^{2}}(|r^{2}-r^{2}_{2m}|)^{3/2},
\end{equation}
and consequently, it is more regular near the point $r=r_{min}$ than
(3.18) considered in Section 3. Indeed, we see from (4.21)
that $g(r)$ and $g'(r)$ are regular in $r=r_{2m}$ though the second
derivative is still singular. It means that the behaviour of
the geodesics is  regular near this point (since only the first
derivatives of the metric enter into equations for geodesics) but
the 4D curvature is singular:  $R^{(4)} \rightarrow + \infty$ if $r
\rightarrow r_{2m}$.  For large $r>>r_{2m}$ the potential (4.19)
asymptotically coincides with (3.17):
\begin{equation} U(r) \approx {r
\over (r^{2}-16 \kappa)^{1/2}} \end{equation} obtained in the previous
section. Hence, we have the same asymptote (3.20), (3.21) for the metric
$g(r)$ and the curvature $R^{(4)}(r)$ for large $r>>r_{2m}$.
Formally, there exists an extension of the metric beyond the point
$r=r_{2m}$.  Indeed, we may again consider the variable $r=r_{2m}
\cosh{x}$.  Then, we come to the same picture as in Section 3. with two
sheets sewed on the hypersurface $r=r_{2m}$. The only
difference from Section 3. is that the singularity at the minimal
radius $r=r_{2m}$ is more mild now.

The potential (4.19) determines also a non-trivial metric defined in the
compact region $0<r \leq r_{1m}$. One can see that this metric describes
the space-time which
has singularities (of curvature) at the points $r=0$ and $r=r_{1m}$.
The curvature near  $r=0$ has the form: $R^{(4)}_{}(r) \approx {2 \over
r^{2}} (1 - e^{-{4 \over B}})$, while the behaviour of the metric near
$r=r_{1m}$ is similar to that  near  $r=r_{2m}$. This
space-time has no asymptotically flat region and is not connected with
the space-time defined for $r \geq r_{2m}$. So it is not observable for
any observer staying at $r>r_{2m}$. The physical meaning of such a
space-time is not clear for us.

Thus, the general picture in the case when the Faddeev-Popov ghosts are
taken into account ($A=0, B \neq 0$) mainly repeats the picture
considered in the previous section.  We may also conclude that the
influence of the ghosts is in smoothering of the
singularity at the minimal radius $r_{min}$.

\bigskip

{\sl 4.2  $A \neq 0$ }

The expression for $\beta^{T}$ is covariant with respect to the target
metric $G_{ij}$. To find $\beta^{T}$ we use the fact that $G_{ij}$ is
flat and consequently it can be reduced to the standard diagonal
form by means of the  coordinate transformation in the target space.
Following [14], the target metric (4.5) \begin{equation}
ds^{2}_{targ}= {1 \over \psi}(1 -{B \over 4\psi}) d\psi^{2} + 4(1- {B
\over 4\psi}) d \psi d \sigma + A d \sigma^{2} \end{equation} by
introducing the new target "coordinates" $(\omega, \chi)$
\begin{equation} \omega^{2}=4\psi ; \ \chi ={1 \over 2} \sigma +{\psi
\over A} -{B \over 4A} \ln 4\psi \end{equation} can be reduced to
the form
\begin{equation} ds^{2}_{targ}=4A d \chi^{2} -{1 \over A} {1
\over \omega^{2}}({\omega^{2} } -B)({\omega^{2} } -A-B) d \omega^{2}
\end{equation}
Note again that $B>0,B+A>0$ for any $N$.

Let $A+B > B$. We see that the second term in (4.25) is positive if $\omega$
lies in the intervals $I_{1}= (0, B)$ or $I_{3}= (A+B, + \infty)$ and it
changes the sign
if $\omega$ lies in $I_{2}=(B,A+B)$.

Let us consider the new  variable $\Omega$
\begin{equation}
\Omega =  \int\limits_{}^{\omega}
{dy \over y} \sqrt{ \pm (y^{2} -B) (y^{2} -A-B)},
\end{equation}
where one must take sign $(+)$ for the intervals $I_{1}, \ I_{3}$ and
$(-)$ for the interval $I_{2}$.

In the new coordinates $(\chi,\Omega)$, the metric (4.25) is diagonal
\begin{equation}
ds^{2}_{targ}=4A d\chi^{2} -( \pm {1 \over A}) d \Omega^{2}.
\end{equation}
Then, we get for the tachyon $T$ in terms of the new variables
\begin{eqnarray}
&&T=-{1 \over 4\kappa} U(\psi)e^{2\sigma} =-{1 \over 4\kappa}U(\omega)
e^{-{\omega^{2}
\over A}} \omega^{2B \over A} e^{4\chi} \nonumber \\
&&\equiv -{1 \over 4\kappa}
\hat{U}(\Omega) e^{4\chi}
\end{eqnarray}
 For $\beta^{T}$  we obtain
\begin{equation}
\beta^{T}=-[{1 \over 4A} \partial_{\chi}^{2} T - (\pm A)
\partial_{\Omega}^{2}T] .
\end{equation}
Inserting (4.28) into (4.29) we obtain the $\beta$-function for the
potential $\hat{U}$
\begin{equation} \beta^{\hat{U}}=(\pm A)
\partial_{\Omega}^{2} \hat{U} -{4 \over A} \hat{U} \end{equation} where
one must take sign $(+)$ for  the intervals $I_{1}, I_{3}$ and $(-)$
for $I_{2}$.

As before, we consider now the renormalization group equation
\begin{equation}
\partial_{t} \hat{U} = ( \pm ) A \partial_{\Omega}^{2} \hat{U}
 -{4 \over A} \hat{U},
\end{equation}
where $t=ln{\mu \over \mu_{0}}$.
Assuming that for $t=0$ the potential $U(\omega,t)$ coincides with the
"bare" one:  $U(\omega,t=0)=1$, we get that eq.(4.31) should be
supplied with the initial condition
\begin{equation}
\hat{U}|_{t=0}= \phi (\Omega) \equiv
e^{-{\omega^{2}(\Omega) \over A}} (\omega (\Omega))^{{2B \over A}}
\end{equation}
where $\omega (\Omega)$ is the inverse function to (4.26).

We see that $\hat{U}$ satisfies the differential equations of
different type in various intervals. Let us assume that $A>0 \ (N>24)$.
Then (4.31) is the standard heat equation being considered in
the intervals $I_{1}, I_{3}$, while it is the heat equation "with
decreasing time" in $I_{2}$.  The problem is that the "decreasing time"
heat equation is known to be non-correct. Solving it formally by means
of the Fourier transform  in the interval $I_{2}$ we get:
\begin{equation}
\hat{U} (\Omega, t) = e^{- 4t/A} \sum_{k=1}^{+ \infty} a_{k} e^{A({\pi
k \over \Lambda})^{2} t} \sin {\pi k \over \Lambda} \Omega ,
\end{equation} where we assumed that in $I_{2}$ the variable $\Omega$
changes in the interval $(-\Lambda , 0)$; $a_{k}=(\phi( \Omega), \sin
{\pi k \over \Lambda \Omega }) $ are the Fourier coefficients of
the initial condition (4.32). In order the sum (4.33) to be convergent
for any finite $t$, the coefficients $a_{k}$ must decrease faster
than the exponent. It is not  obviously the case for the initial
condition of the type (4.32).  So the solution of (4.31) in the
interval $I_{2}$ does not exist at least within the quadratically
integrable functions (probably (4.31) can be solved in the class of
distributions).

If $A<0 \ (N<24)$ eq.(4.31) is the standard heat equation only in the finite
interval $I_{2}=(A+B,B)$, while it is again of the "decreasing time" type  in
semi-infinite
intervals $I_{1},I_{3}$ with the same problems concerning the solution
as above (the formal solution takes the same form as (4.33)
changing the sum $\sum_{k}^{}$ by the integral $\int\limits_{0}^{+
\infty} dk$). We have no definite idea about the possible
physical interpretation of the solution valid only in the interval
$-\Lambda < \Omega  < 0$ or equivalently $2\kappa (A+B) < r^{2} <
2\kappa B$. So we will consider only the case of positive $A$.

Let $A>0$ and consider eq.(4.31) in the interval $I_{3}$. We may
choose the integration constant in $I_{3}$ to arrange that $\Omega
(A+B)=0$. Then, $\Omega (\omega)  $ determined by (4.26) varies in the
interval $(0,+ \infty)$
\begin{equation} \Omega = {1 \over 2}
\sqrt{(\omega^{2}-B)(\omega^{2}-A-B)} - {A+2B \over 2}
\ln{\sqrt{\omega^{2}-B} + \sqrt{\omega^{2}-A-B} \over \sqrt{A}}
\end{equation}

$$- {\sqrt{B(A+B)} \over 2} \ln{2B(A+B)-(A+2B)\omega^{2} +2\sqrt{B(A+B)}
\sqrt{(\omega^{2}-B)(\omega^{2}-A-B)} \over -A\omega^{2}}
$$

Thus, the function $\hat{U}$ is defined in the region $\{ t \geq 0,
\Omega \geq 0 \}$.  Hence  we should have  an appropriate boundary
condition at $\Omega =0$. In fact, the results (the behaviour of
 $g(r)$ and $R^{(4)}(r)$ and possibility of extension on
$(-)$-sheet) are not changed if we take an arbitrary condition:
$\hat{U}|_{\Omega =0} = \mu (t)$.  However, for simplicity, we choose the
zero boundary condition:  \begin{equation} \hat{U}|_{\Omega=0} =0
\end{equation}
For the chosen initial and boundary conditions  eq.(4.31) has
the solution:
\begin{equation}
\hat{U} (\Omega,t)={ e^{-{4t \over A}} \over \sqrt{At} } {1 \over 2\sqrt{\pi}}
\int\limits_{0}^{+ \infty}[e^{-{(\Omega - \xi)^{2} \over 4At}} -
e^{-{(\Omega + \xi)^{2} \over 4At}} ] \phi (\xi) d\xi ,
\end{equation}
where $\xi$ and $\omega(\xi)$ are related by
\begin{equation}
\xi=  \int\limits_{\sqrt{A+B}}^{\omega(\xi)} {dy \over y} \sqrt{({y^{2} } -B)
({y^{2} } -A-B)} .
\end{equation}
We are interested in the potential $U(\omega,t)= e^{\omega^{2} \over A}
\omega^{-2B \over A} \hat{U} (\Omega, t)$, where $\omega$ and
$\Omega$ are related by (4.26). Note that the $t$-dependence of
the solution (4.36) can be again absorbed into the  redefinition of the
constants $A,B,\kappa$. Indeed, let us consider the "renormalized"
constants $A_{}^{*}={A \over t}, B_{}^{*} ={B \over t},
\kappa_{}^{*}=\kappa t$.  Note that $\omega^{2} \equiv {r^{2} \over
2\kappa} = t \omega_{}^{2} (\kappa_{}^{*}), \ \Omega (A,B,\kappa)= t
\Omega(A_{}^{*}, B_{}^{*},\kappa_{}^{*})$. We assume that
$A^{*},B^{*},\kappa^{*}$ coincide with their "observable"values
($A^{*}={N-24 \over 12\pi}, \ B^{*}={24 \over 12\pi}$).  Then, in terms
of the new constants we obtain the  potential $U(\omega)$
\begin{equation} U(\omega)={1 \over \sqrt{A} }e^{-{4 \over A}}{1 \over
2\sqrt{\pi}} \int\limits_{0}^{+ \infty}[e^{-{(\Omega - \xi)^{2} \over
4A}} - e^{-{(\Omega + \xi)^{2} \over 4A}} ] ({\phi(\xi) \over \phi
(\Omega)}) d\xi, \end{equation}
where $\omega^{2}={r^{2} \over 2\kappa^{}}$ and  we have omitted $*$.
The potential $U(r)$ for different $(A,B)$ is plotted in Fig.1.

We interpret  the 4D metric (2.14)
\begin{equation}
g(r)={-2M \over r} + {1 \over r} \int\limits_{r_{min}}^{r} U(\rho)d\rho
\end{equation}
where $r_{min}= \sqrt{2\kappa(A+B)}$,
with the potential $U(r)$ in the form (4.38) as the Schwarzschild
metric deformed due to the quantum spherically symmetric excitations of the
ghosts and matter fields.
The role of quantum fluctuations of the field
$f$ is  only in vacuum polarization around the gravitating body which
leads to the right hand side of the Einstein equations (2.19) in the form
(2.20). A different situation would happen when the collapse of  the
$f$-field impulse is considered [6-7].  Then,  we would have to take into
account the back-reaction of the Hawking radiation that needs the
static solutions of equations obtained by varying the quantum
action (4.3). We do not consider it here.

Analyzing the metric (4.39) we note that here  the minimal
distance $r_{min}= \sqrt{2\kappa(A+B)}$ again appears. The
expression (4.39) is valid  only for $r \geq r_{min}$. The essential
difference of the metric (4.39) with the potential (4.38) from that of
(3.19) or (4.21) is that it is more regular near the point $r=r_{min}$. To
see this, note that the solution of the heat equation $\hat{U}(\Omega)$
(4.36) in the vicinity of the point $r=r_{min}$  takes
the form $\hat{U}(\Omega)=c \Omega$ ($0< \Omega <<1$), where $c>0$ is an
irrelevant constant.  Then, we have for $U(\omega)$:
\begin{equation} U(\omega)=c
e^{\omega^{2} \over A} \omega^{-{2B \over A}} \Omega(\omega) ,
\end{equation}
where $\omega^{2}={r^{2} \over 2\kappa}$. From
(4.40) and (4.34) we obtain for $r \approx r_{min}$ that

$$
U(r) \approx c'
(r^{2} - r^{2}_{min})^{3/2} .
$$
Thus, the metric function $g(r)$ (4.39) near  $r=r_{min}$ can be
approximately written in the form
\begin{equation}
g(r) \approx 1-{2M \over r} + {C \over r} (r^{2} - r_{min}^{2})^{5/2},
\end{equation}
where $C$ is some constant.
So we obtain that $g(r)$ and its derivatives $g'$ and $g''$ are regular
at $r=r_{min}$ and only the third derivative diverges $g'''(r_{min})=+
\infty$. Consequently, the 4D scalar curvature (2.17) takes finite value
at $r=r_{min}$ though its derivative diverges
 $R'^{(4)}(r_{min})=-\infty$.  This non-analyticity  implies a rather
 mild singularity which does not affect the behaviour of the geodesics.
The latter is regular near $r=r_{min}$ that implies an analytic extension
of the space-time beyond the hypersurface $r=r_{min}$ (in the opposite case
we would obtain the manifold with the boundary at $r=r_{min}$ that
seems to be unsatisfactory). As we have discussed above, it cannot be
the extension to the small radius $r<r_{min}$ since we have no
solution of our equations in this region.  To construct such an
extension, we note that the variable $\Omega$ was introduced in such a
way that its differential squared $(d\Omega)^{2}$ gives the second term
in (4.25).  It is clear that for fixed $\omega$ both $\Omega$ (4.26)
and $- \Omega$ satisfy this condition. We use this fact to consider
$\Omega (\omega)$ as a two-fold function in the interval $I_{3}$.
Then, we can continue the above obtained expression and valid for
$\Omega >0$ to the  negative values of $\Omega$.

We see from (4.36) that $\hat{U}(\Omega)$ continued onto the interval
$(- \infty, + \infty)$ is an  odd function:
$\hat{U}(-\Omega)=-\hat{U}(\Omega)$. As a result, for the fixed $r$ the
metric function $g(r)$ takes two values  \begin{equation}
g^{(\pm)}(r)={-2M \over r} \pm {1 \over r} \int\limits_{r_{min}}^{r}
U(\rho)d\rho ,\end{equation} where $U(\rho)$ is given by (4.38). The
corresponding expression for the 4D scalar curvature is
\begin{equation}
R^{(4)}_{(\pm)}= {2 \over \kappa  r^{2}} (1 \mp U(r)) \mp {U' \over
\kappa  r} .
\end{equation}
Thus, we obtain the same picture as considered in Section 3.
The metric (4.42) describes the 4D space-time with two sheets ($g^{( \pm)}$
is the metric on $(\pm)$-sheet) which are glued on the hypersurface of
constant radial coordinate $r=r_{min}$. The functions $g^{(+)}(r)$,
$g^{(-)}(r)$ and their first and second derivatives are regular and
sewed continuously at $r=r_{min}$ and only the third derivatives
diverge $g^{'''(\pm)}(r_{min})=\pm \infty$. As a result, we see that one
sheet is the extension of the other\footnotemark\
\addtocounter{footnote}{0}\footnotetext{One can see this transparently by
using the  variable $x$: $r=r_{min} \cosh{x}$. Then, $\Omega (-x)=- \Omega
(x)$ and the $(-)$-sheet corresponds to $x<0$. Metric is extended
continuously into this region.}: the geodesics of one sheet are not
ended at $r=r_{min}$ but continuously extended to the other sheet.  One
can see from (4.38) that $U(r)>0$ and hence $g^{(-)}(r) <0$ while
$g^{(+)}(r)$ has zero at the single point $r_{h}$ which is a solution of the
equation $2M= \int\limits_{r_{min}}^{r_{h}} U(\rho)d\rho$. To see the
behavior of (4.42) at large distances $(r>>r_{min})$ it is useful to
note that the potential $U(\omega)$ (4.38) for $\omega^{2}>>A+B$
asymptotically coincides with (4.19), (3.19):
\begin{equation} U(r) \sim {r
\over (r^{2}- 16 \kappa)^{1/2}}
\end{equation}
and consequently, the metric
$g^{(\pm)}(r)$ (4.42) behaves for large $r>>r_{min}$ as follows:
\begin{equation}
g^{(\pm)} (r)  \sim 1-{2M \over r} \pm {(r^{2}-16 \kappa)^{1/2} \over r}
\end{equation}
Both sheets are asymptotically flat. However, the flatness is reached
in a different way that one can see from the scalar
curvature (4.43) at large distances:
\begin{equation}
R^{(4)}_{(+)} \sim {2 \over  r^{2}} ({a \over r})^{4} \ ; \
R^{(4)}_{(-)} \sim {4 \over  r^{2}}.
\end{equation}
This coincides with what we have obtained in Section 3. (3.20), (3.24).
The plot of $R^{(4)}_{(+)}(r)$ is shown in Fig.2. We see that the
space-time has a horizon located on the $(+)$-sheet at $r=r_{h}> r_{min}$
and the whole $(-)$-sheet is behind this horizon from the point of view
of an observer staying at $r>r_{h}$ on $(+)$-sheet. The topology of
$t=const$ slice of the space-time is shown in Fig3. The Penrose diagram
of the space-time  can be seen in Fig.4.

It is worth noting that the resulting space-time does not much differ from that
of
the black hole with the internal de Sitter space instead of singularity. Such
space-times have been earlier considered in [15] where it has been shown
that such a picture may occur under the condition that limiting
curvature exists and the Schwarzschild singularity does not arise. In
case of the two-dimensional black holes such solutions were considered in
a number of papers [16].

If now $\omega$ lies in the interval $I_{1}=(0,B)$ (or equivalently the
radius $r$ lies in $(0, \sqrt{2\kappa B})$), the function
$\Omega(\omega)$ (4.26) takes the form  \begin{equation}
\Omega(\omega)=\int\limits_{\omega}^{\sqrt{B}} {dy \over y}
\sqrt{(B-y^{2})(A+B-y^{2})} ,
\end{equation}
where we have chosen the integration constant so that $\Omega(B)=0$. For
$0<\omega< \sqrt{B}$ we get that $\Omega(\omega) \geq 0$ and
 $\Omega(0)=+ \infty$.  The solution of eq.(4.31)
(with sign $(+)$) takes the form (4.36). For the potential $U(\omega)$ we
get the expression (4.38) where now $\Omega(\omega)$ (and
$\omega(\xi)$) is given by (4.47). Proceed analogously, we also can
consider here the $(\pm)$-sheets with the metric $g^{\pm}(r)$,
respectively:
\begin{equation} g^{(\pm)}(r)=-{2M \over r} \mp
{1 \over r}\int\limits_{r}^{r_{0}} U(\rho)d\rho ,\end{equation}
where $r_{0}=\sqrt{2\kappa B}$. The curvature is given by (4.43). The
space-time is regular at $r=r_{0}$ and both sheets are sewed at $r=r_{0}$
continuously in the same manner as above.
However, we again obtain the singularity at $r=0$ since
the 4D scalar curvature in the limit $r \rightarrow 0$ tends to
\begin{equation}
R^{(4)}_{(\pm)} \sim {2 \over r^{2}} [1 \mp e^{-{4 \over
A+B}}]
\end{equation}

Thus, for $A>0$ the general picture is similar to that we have obtained for
$A=0$: the
solution of eq.(4.32) describes  two non-connected space-times.  The first
one , located at $r \geq r_{min}$, is free from the singularities  and
is asymptotically flat, while the second, located in the region $0
\leq r \leq r_{0}$, is singular at $r=0$.  However, only the first
space-time has an obvious physical meaning and realizes our
preliminary assumption that quantum corrections might lead to drastic
deformation of the Schwarzschild solution and avoid space-time
singularity.

\section{Conclusion}
\setcounter{equation}0

We have studied  the problem of deformation of the Schwarzschild
solution due to quantum corrections in the approximation when only
spherically symmetric excitations are taken into account.

One of our predictions concerns the behavior of the metric function
$g(r)$ and the corresponding curvature $R^{(4)}(r)$ outside the
gravitating body with the mass $M$ at distances much  larger than
the Planck scale, $r>>a=4\sqrt{\kappa}$:
\begin{equation} g(r)
\approx 1-{2M \over r} - {1 \over 2}({a \over r})^{2} -{1 \over 8}
({a \over r})^{4} \ ; \ R^{(4)}(r) \approx
{2 \over r^{2}} ( {a \over r})^{4} .\end{equation}
It is worth noting
that this expression is rather universal and does not depend on
whether gravitational ghosts and matter contributions are included
in the consideration or not.  One can see from (5.1) that the
space-time outside the gravitating body is no more Ricci flat
as it follows from the classical Einstein equations, though the scalar
curvature $R^{(4)}(r)$ rapidly tends to zero and becomes too negligible
to be observed in present gravitational experiments.

The other important point is the behavior of the space-time near the
Schwarzschild singularity. We have  shown that quantum corrections
lead to the shift of the singularity at $r=0$ to the finite
distance $r_{min} \sim r_{pl}$ and make it smoother.
The scalar curvature $R^{(4)}(r)$ takes the finite value at
$r=r_{min}$, so the space-time looks regularly  near this minimal
radius and allows the analytic extension beyond it. The complete
space-time is free from singularities and consists of two asymptotically
flat sheets glued on hypersurface of constant radial parameter
$r=r_{min}$, so that  one sheet is behind the horizon with respect
to an observer staying on the other sheet. This is the result of the
deformation of the Kruskal extension of the classical Schwarzschild metric
due to quantum corrections. We see that these corrections indeed make
the singular classical space-time more regular as it was
originally assumed.

The method developed can be applied to the study of the other known
classically singular solutions of general relativity: the
Reissner-Nordstrom and cosmological ones.  This work is in progress.

The next problem of interest is how the  corrections found may change
the gravitational collapse. It is known that the collapse in the classical
Einstein gravity  ended by formation of the singularity. Probably our
results mean that the real singularity is not formed and at the
end of gravitational collapse one obtains the regular
space-time.  However, at the present stage we can not make any
definite  conclusion since the Hawking radiation and its
back-reaction were not taken into account. The previous study of 2D
black holes tells us that the space-time points such as $r=r_{min}$ at
which the $D=2$ $\sigma$-model becomes degenerate (see eq.(4.6)) are the
possible places for new singularities to be formed [17]. This
problem needs further investigation.

\newpage

{\bf Figure Captions} \\
Fi1. 1. The shape of the "dilaton" potential $U(r,A,B)$ for: a). $A=0.1, \
B=1$;
b). $A=0.3, \ B=1$; c). $A=1, \ B=1$.

Fig. 2. The shape of the 4D scalar curvature $R^{(4)}(r)$ induced by quantum
corrections  for $A=0.1, \ B=1$.

Fig.3. The $t=const$ slice of the extended  Schwarzschild
space-time deformed by quantum corrections. It consists of two asymptotically
flat $(\pm)$-sheets glued on
the hypersurface of constant radial parameter $r=r_{min}$. The event horizon
is located on the hypersurface $r=r_{h}$ of the $(+)$-sheet.

Fig.4. The Penrose diagram of the space-time deformed by quantum corrections
for $A,B>0$. The asymptotically flat $(+)$-region has the same  causal
properties
as the classical Schwarzschild solution.
It is analytically extended beyond the hypersurface $r=r_{min}$
to the other asymptotically flat $(-)$-region, so that the complete space-time
is free from singularity.


\begin{thebibliography} \\
\bibitem{1d}  R.Penrose, Phys.Rev.Lett. {\bf 14} (1965), 57;
              S.Hawking, Proc.R.Soc. (London) {\bf A300} (1967), 182;
              S.Hawking and R.Penrose, Proc.R.Soc. (London) {\bf A314} (1970),
              529.

\bibitem{2d}  M.J.Duff, Phys.Rev. {\bf D9} (1974), 1837.

\bibitem{3d}  V.P.Frolov, G.A.Vilkovisky, Phys.Lett. {\bf B106} (1981), 307.

\bibitem{4d}  G.'t.Hooft, Phys.Lett. {\bf B198} (1987), 61; Nucl.Phys. {\bf
B304}
              (1988), 867; {\bf B335} (1990), 138.

\bibitem{5d}  H.Verlinde and E.Verlinde, Nucl.Phys. {\bf B371} (1992), 246.

\bibitem{6d}  G.Mandal, A.Sengupta and S.Wadia, Mod.Phys.Lett.
                           {\bf A6} (1991), 1685;
              E.Witten, Phys.Rev. {\bf D44} (1991), 314;
              C.G.Callan, S.B.Giddings, J.A.Harvey and A.Strominger,
              Phys.Rev. {\bf D45} (1992), R1005;
              T.Banks, A.Dabholkar, M.R.Douglas and M.O.'Loughlin,
              Phys.Rev. {\bf D45} (1992), 3607.

\bibitem{7d}  J.A.Harvey and A.Strominger, {\it "Quantum Aspects of Black
Holes"},
              Enrico Fermi Institute Preprint (1992), hep-th/9209055;
              S.B.Giddings, {\it "Toy Model for Black Hole Evaporation"},
              UCSBTH-92-36, hep-th/9209113.

\bibitem{8d}  M.Guigan, C.R.Nappi and S.A.Yost, {\it "Charged Black Holes In
              Two Dimensional String Theory"}, IASSNS-HEP-91/57;
              O.Lechtenfeld and C.Nappi, {\it "Dilaton Gravity and No hair
              Theorem in Two Dimensions" }, IASSNS-HEP-92/22;
              D.A.Lowe, Phys.Rev. {\bf D47} (1993), 2446.

\bibitem{9d}  S.P.Trivedi,{\it "Semiclassical Extremal Black Holes"},
              CALT-68-1833 (1992);
              A.Strominger and S.P.Trivedi, {\it "Information Consumption by
              Reissner-Nordstrom Black Holes"}, NSFITP-93-15/CALT-68-1851
              (1993).

\bibitem{10d} T.Banks, M.O.'Loughlin, Nucl.Phys. {\bf B362} (1991), 649.

\bibitem{11d} V.P.Frolov, Phys.Rev. {\bf D46} (1992), 5383.

\bibitem{12d} J.G.Russo, A.A.Tseytlin, Nucl.Phys. {\bf B382} (1992), 259.

\bibitem{13d} A.Strominger, Phys.Rev. {\bf D46} (1992), 4396.

\bibitem{14d} A.Bilal and C.Callan, Nucl.Phys. {\bf B394} (1993), 73.

\bibitem{15d} V.P.Frolov, M.Markov and V.Mukhanov, Phys.Rev. {\bf D41} (1990),
383;
              E.Fahri and A.Guth, Phys.Lett. {\bf B183} (1987), 149;
              A.Linde, Nucl.Phys. {\bf B372} (1992), 421;
              D.Morgan, Phys.Rev. {\bf D43} (1991), 3144.

\bibitem{16d} T.Banks and M.O.'Loughlin, {\it "Non-singular Lagrangians for
              Two-dimensional Black Holes" }, RU-92-63, hep-th/9212136;
              D.A.Lowe and M.O.'Loughlin, {\it "Non-singular Black Hole
              Evaporation and "Stable Remnants" " }, RU-93-12/PUPT-1399,
              hep-th/93-05125;
              M.Trodden, V.F.Mukhanov and R.H.Brandenberger, {\it "A
Non-singular
              Two-dimensional Black Holes" }, BROWN-HET-907.

\bibitem{17d} S.W.Hawking and J.M.Stewart, {\it "Naked and Thunderbolt
Singularities
              in Black Hole Evaporation" }, PRINT-92-0362, hep-th/9207105.

\end{thebibliography}
\end{document}